\documentclass[a4paper, 11pt]{article}

\usepackage{jinstpub}

\usepackage{bm}
\usepackage{lineno}
\usepackage{hyperref}
\usepackage{xcolor}
\usepackage{float}
\usepackage{setspace}
\usepackage{overpic}
\usepackage{subfig}

\usepackage{tikz}
\usetikzlibrary{positioning}
\usetikzlibrary{calc}
\definecolor{mylightred}{RGB}{211,79,73}
\definecolor{mydarkred}{RGB}{199,44,38}
\definecolor{mylightgreen}{RGB}{78,153,67}
\definecolor{mydarkgreen}{RGB}{43,129,33}
\definecolor{mylightpurple}{RGB}{150,107,178}
\definecolor{mydarkpurple}{RGB}{126,78,160}
\definecolor{mylightblue}{RGB}{49,101,205}
\definecolor{mydarkblue}{RGB}{20,92,205}
\tikzset{
  juliadot/.style args={#1,#2}{shape=circle,line width=0.03ex,minimum width=0.4ex,fill=#1,draw=#2}
}
\newcommand\julialetter[1]{{\strut\fontfamily{cmss}\bfseries\selectfont{#1}}}

\DeclareRobustCommand\julia{%
\begin{tikzpicture}[baseline=0mm, every node/.style={inner sep=0mm, outer sep=0mm}]
\node[anchor=base]        (j) at (0,0) {\julialetter{\j}};
\node[anchor=base, right=0ex of j] (u) {\julialetter{u}};
\node[anchor=base, right=0ex of u] (l) {\julialetter{l}};
\node[anchor=base, right=0ex of l] (i) {\julialetter{\i}};
\node[anchor=base, right=0ex of i] (a) {\julialetter{a}};
\path let \p1 = (j) in node[juliadot={mylightblue,mydarkblue}] (bluedot) at (\x1+0.02ex,1.4ex) {};
\path let \p1 = (i) in node[juliadot={mylightred,mydarkred}] (reddot) at (\x1,1.4ex) {};
\path let \p1 = (reddot) in node[juliadot={mylightpurple,mydarkpurple}] (purpledot) at (\x1+0.5ex,\y1) {};
\path let \p1 = (reddot) in node[juliadot={mylightgreen,mydarkgreen}] (greendot) at (\x1+0.25ex,\y1+0.42ex) {};
\end{tikzpicture}%
}

\title{Simulation of semiconductor detectors in 3D with {\it SolidStateDetectors.jl}}

\author[a]{I.~Abt,}
\author[a]{F.~Fischer,}
\author[a]{F.~Hagemann,}
\author[a]{L.~Hauertmann,}
\author[a]{O.~Schulz,}
\author[a]{M.~Schuster}
\author[a,1]{and A.J.~Zsigmond\note{Corresponding author.}}
\emailAdd{azsigmon@mpp.mpg.de}
\affiliation[a]{Max-Planck-Institut f\"ur Physik, M\"unchen, Germany}

\abstract{
The open-source software package {\it SolidStateDetectors.jl}
to calculate the fields and simulate the drifts of charge carriers in solid state detectors, especially in large volume high-purity germanium detectors, together with the corresponding pulses, is introduced. The package can perform all calculations in full 3D while it can also make use of detector symmetries. The effect of the surroundings of a detector can also be studied. 
The package is programmed in the user friendly and performance oriented language \julia{}, such that 3D field calculations and drift simulations can be executed efficiently and in parallel.
The package was developed for high-purity germanium detectors, but it can be adjusted by the user to other types of semiconductors.
The verification of the package is shown for an n-type segmented point-contact
germanium detector. Additional features of {\it SolidStateDetectors.jl}, which are under development are listed.
}


\arxivnumber{2104.00109}
\collaboration{\includegraphics[height=20mm]{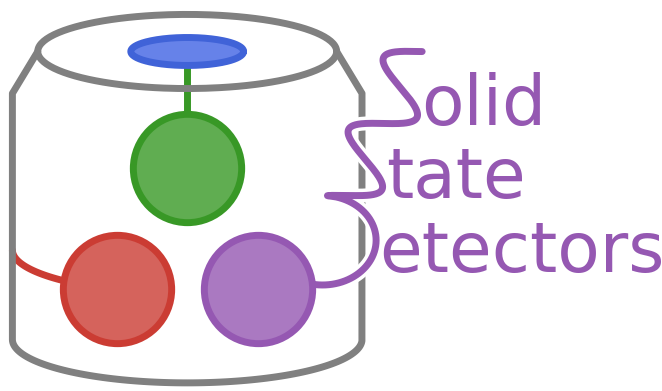}\\[12pt]}

\begin{document}
\maketitle



\section{Introduction}
\label{sec:introduction}

Semiconductor detectors are used widely in industry and in basic as well as applied science.
They play an important role in various nuclear and particle physics experiments.
A special case are high-purity germanium (HPGe) detectors, for which {\it SolidStateDetectors.jl} was originally developed.
Due to the extremely low impurity densities achieved for germanium, typically around $10^{10}$ per cubic centimeter, these detectors can have volumes of several hundred cubic centimeters.
Germanium detectors are known for their good energy resolution, which is in many cases exploited by only reading out one of two electrodes, often called "core".
Cylindrical geometries are common where the core electrode is established either in a central bore hole or as a circular patch on one of the end-plates. In order to minimize the capacitance of such a patch, it is often implemented as small as possible and is called point-contact. Depending on the manufacturer, such detectors are also called "Broad Energy Germanium" (BEGe) detectors.

In AGATA~\cite{Akkoyun:2011ft} and in GRETINA/GRETA~\cite{Fallon:2016lfp, GRETA-CDR}, arrays of segmented cylindrical HPGe detectors are used for spectroscopic measurements using gamma-ray tracking. The outer surface, also called mantle, of an AGATA detector, which is 10\,cm long and 8\,cm in diameter, is divided into 36 electrodes and for a germanium detector, this is considered highly segmented.   
Low-background experiments searching for neutrinoless double beta ($0\nu\beta\beta$) decay~\cite{Ackermann:2012xja, Abgrall:2013rze} or dark matter~\cite{Li:2013fla, Agnese:2014aze, Jiang:2018pic} as well as experiments measuring coherent elastic neutrino-nucleus scattering~\cite{Akimov:2017ade, Belov:2015ufh, Hakenmuller:2019ecb} exploit the radio-purity, the excellent energy resolution and the low energy threshold of point-contact HPGe detectors. These detectors are usually not segmented and the point-contact is the only electrode which is read out.

In modern experiments, not only the energies deposited in the detectors are recorded but also the time development of the charges induced in the read out electrodes, the signals which form pulses.
Pulse shape analysis plays a central role to facilitate position reconstruction in segmented detectors and background rejection in point-contact detectors.
In particular, the rejection of events with more than one energy deposition, i.e.\ multi-site events, in searches for $0\nu\beta\beta$ decay reduces the background induced by gamma-rays by about an order of magnitude~\cite{Agostini:2013jta}, while keeping a high efficiency for single-site events like $0\nu\beta\beta$ decays.
In the case of point-contact or BEGe detectors,
both {\sc Gerda}~\cite{Ackermann:2012xja, Agostini:2013jta} and {\sc Majorana}~\cite{Abgrall:2013rze, Alvis:2019dzt} apply pulse shape discrimination based on the so called $A/E$ parameter, where $A$ is the maximum of the current amplitude and $E$ is the deposited energy.
In an alternative approach, artificial neural networks have been implemented to reject multi-site events recorded in coaxial~\cite{Agostini:2013jta, Abt:2014xta} and BEGe~\cite{Holl:2019xtt} detectors.
Based on the success of {\sc Gerda}~\cite{Agostini:2020xta} and {\sc Majorana}~\cite{Alvis:2019sil} in achieving good sensitivity to $0\nu\beta\beta$ decay, the next generation experiment LEGEND~\cite{Abgrall:2017syy} will, in its first phase, operate about 200\,kg of HPGe detectors under low-background conditions.
In its second phase, LEGEND will be upgraded to 1000\,kg.

A precise simulation of the detector pulses is essential to fully understand the effects of pulse-shape discrimination techniques and their efficiencies, and to provide clean training samples for neural networks.
One of the technical applications of simulation is that, during the production of detectors for LEGEND, the cutting of the available crystals can be optimized to
make the production of detectors with maximal masses and still
reasonably small bias voltages possible.
This optimization has to take the active impurity profiles of the crystals into account.
Thus, a simulation package is needed, where not only the geometry but also any kind of impurity profile can easily be implemented. In addition, the electric field calculation including the handling of undepleted regions has to be fast, such that many scenarios can be studied in a reasonable amount of time.
However, the package is not an engineering tool.
It does not provide any checks whether a detector is technically feasible.
The package provides the functionality to predict the response of a given detector configuration in a given environment.

Previously published packages to simulate the pulses of HPGe detectors do not take the detector environment into account.
In addition, they are either not open source~\cite{Agostini:2010rd, Bruyneel:2016zih} or difficult to extend~\cite{Abt:2010ax, Radford:siggen}.
The new open-source pulse-shape simulation package, {\it SolidStateDetectors.jl}, fulfills all requirements of LEGEND related research activities and provides a modular software environment that can easily be extended by the user for other applications like different
HPGe detector geometries or special silicon detectors.


\section{Detector geometries and simulation procedure}
\label{sec:overview}

The user defines the "world", including the geometry of the detector and, optionally, its surroundings as well as the electric boundary conditions in structured text configuration files.
Constructive solid geometry (CSG) is used to define different shapes and to construct all objects.
The coordinate system, either cylindrical or Cartesian, is also defined in the configuration file as well as the boundaries of the world.
{\it SolidStateDetectors.jl} provides example configuration files for selected common HPGe detectors, including a segmented true coaxial~\cite{Abt:2007rf}, a segmented BEGe~\cite{Abt:2018yvd} and an inverted coaxial~\cite{Cooper:201125} detector\footnote{The latter is the baseline design for the LEGEND experiment because such detectors have excellent pulse-shape discrimination properties and large masses.}.
It also provides an interface to read in configuration files from {\sc Majorana} Siggen~\cite{Radford:siggen}.

The simulation can be divided into two main parts: 1) the calculation of the static electric properties of a given experimental setup and 2) the drift of charge carriers inducing signals on the electrodes. 
Both of these are implemented in the programming language  \julia{} in a modular manner. 
The \julia{} language has a high-level syntax leading to an easy to use environment but it is still designed for high performance~\cite{Julialang:2017}.

The electric field is calculated once for a given detector geometry at the beginning of the simulation, based on the electric properties of the system.
The main properties are the impurity density profile of the crystal and the fixed potentials of the electrodes.
In addition, the user can define volumes with fixed charges on the detectors and surrounding materials at fixed electrical potentials, like grounded holding structures.
Similarly, the weighting potentials, which are used to determine the signals induced on the electrodes by the charge carriers, are calculated once for each detector configuration. 

For an event-by-event simulation, the interaction of radiation with the detector and surrounding material can be simulated with a dedicated software package like 
{\sc Geant4}~\cite{Agostinelli:2002hh}. 
The positions of the individual energy depositions (hits) of each event are used as input to the pulse generation. 
{\it SolidStateDetectors.jl} provides flexible clustering of the input hits through the service package {\it Clustering.jl}; this can be performed when the hits are read in.
For each hit or cluster, the induced charge carriers are drifted in user defined time steps, default is 1\,ns, according to the previously calculated electric field and the implemented drift velocity model. 
At each step, the induced signals on the electrodes are computed using the weighting potentials. 


\section{Electric field and weighting potentials}
\label{sec:field}

The electric field, $\bm{\mathcal{E}}(\mathbf{r})$, is calculated numerically as the first derivative of the electric potential, $\Phi(\mathbf{r})$, which itself is calculated numerically according to Gauss' law:
\begin{equation} \label{eq:gauss_law}
    \nabla \left( \epsilon_0 \epsilon_{r}(\mathbf{r}) \nabla \Phi(\mathbf{r}) \right) = - \rho(\mathbf{r}) \text{\,,}
\end{equation}
where $\epsilon_0$ is the vacuum permittivity, $\epsilon_{r}(\mathbf{r})$ and $\rho(\mathbf{r})$ describe the relative permittivity and the charge density as functions of the position $\mathbf{r}$.
The input on $\epsilon_{r}(\mathbf{r})$, $\rho(\mathbf{r})$ and the boundary conditions is specified by the user in the configuration file. 
The charge density inside a detector has two different components: $\rho(\mathbf{r}) = \rho_{\mathrm{imp}}(\mathbf{r}) + \rho_{\mathrm{fix}}(\mathbf{r})$, where $\rho_{\mathrm{imp}}(\mathbf{r})$ is due to the impurity density of the semiconductor and $\rho_{\mathrm{fix}}(\mathbf{r})$ is an optional fixed contribution, e.g.\ a charged surface layer.

The numerical calculation is performed using a successive over-relaxation (SOR) algorithm on an adaptive grid with red-black division for parallelization. 
The two nearest neighbors in each dimension are used to numerically update the potential, $\Phi_i$, on the grid point $i$.
At the boundary of the world, a prescription how to update the potential of the point $j$, which is located on the boundary, is needed.
An extra point $j+1$ is introduced, for which the user can choose for each axis individually the prescription: reflecting ($\Phi_{j+1}=\Phi_{j-1}$), periodic ($\Phi_{j+1}=\Phi_{1}$),
fixed ($\Phi_{j+1}\,=$\,constant, normally $\Phi_{j+1}=0$),
decaying ($\Phi_{j+1}=\Phi_{j} d_j/d_{j+1}$, where $d_j$ ($d_{j+1}$) is the distance between point $j$ and the origin in this dimension).
Reflective or periodic are good choices if the symmetries of a system are to be used.
Periodic is especially useful for the $\phi$-direction, where it allows to calculate the fields of rotationally symmetric detectors in 2D instead of in full 3D.

At the beginning of the calculation, a very coarse grid is initiated. 
Once the SOR algorithm has converged, i.e.\ changes are below a user defined value, grid points are inserted in a dimension if the difference in potential for two neighboring points is above a user defined threshold . 
The default number of such grid refinements is three.
However, the user can set the number to a higher value to ensure that the targeted precision given by the threshold is achieved.
The gradual adaptive refinement of the grid avoids large gradients between grid points and ensures fast convergence of the calculation as it does not create a large number of points in volumes of moderately changing potentials. 
The electric field at a given position between grid points is evaluated by linear interpolation.

There is an option that during an iteration within the SOR algorithm, grid points are marked as undepleted.
As the net charge carrier density is zero in the undepleted regions of a semiconductor, 
$\rho_{\mathrm{imp}}(\mathbf{r})$ is set to zero for such grid points.
This feature allows studying the development of the depletion zone with increasing bias voltage for different detector geometries and impurity distributions in order to optimize detector configurations. 
Especially point-contact detectors can, depending on the impurity density profile, develop volumes which cannot be depleted. After a crystal is produced, it is essential to cut it such that this problem is avoided. Simulation is the best guide for this. 

The weighting potential, $\Phi^w_i(\mathbf{r})$, determines the size of the signal induced on the electrode $i$ as a function of $\mathbf{r}$.
It is calculated by solving
\begin{equation}
    \nabla \left( \epsilon_{r}(\mathbf{r}) \nabla \Phi^w_i(\mathbf{r}) \right) = 0
\end{equation}
with the boundary conditions $\Phi^w_i(\mathbf{r}_i)=1$ and  $\Phi^w_i(\mathbf{r}_j)=0$ for all $\mathbf{r}_i$ on electrode $i$ and all $\mathbf{r}_j$ on electrodes $j \neq i$. 
The same SOR algorithm as used to calculate the electric potential is used to calculate the weighting potentials. 

\begin{figure}
    \centering
    \vspace{0.5cm}
    \subfloat[Top view.]{\begin{overpic}[width = 0.50 \textwidth,,tics=10]
    {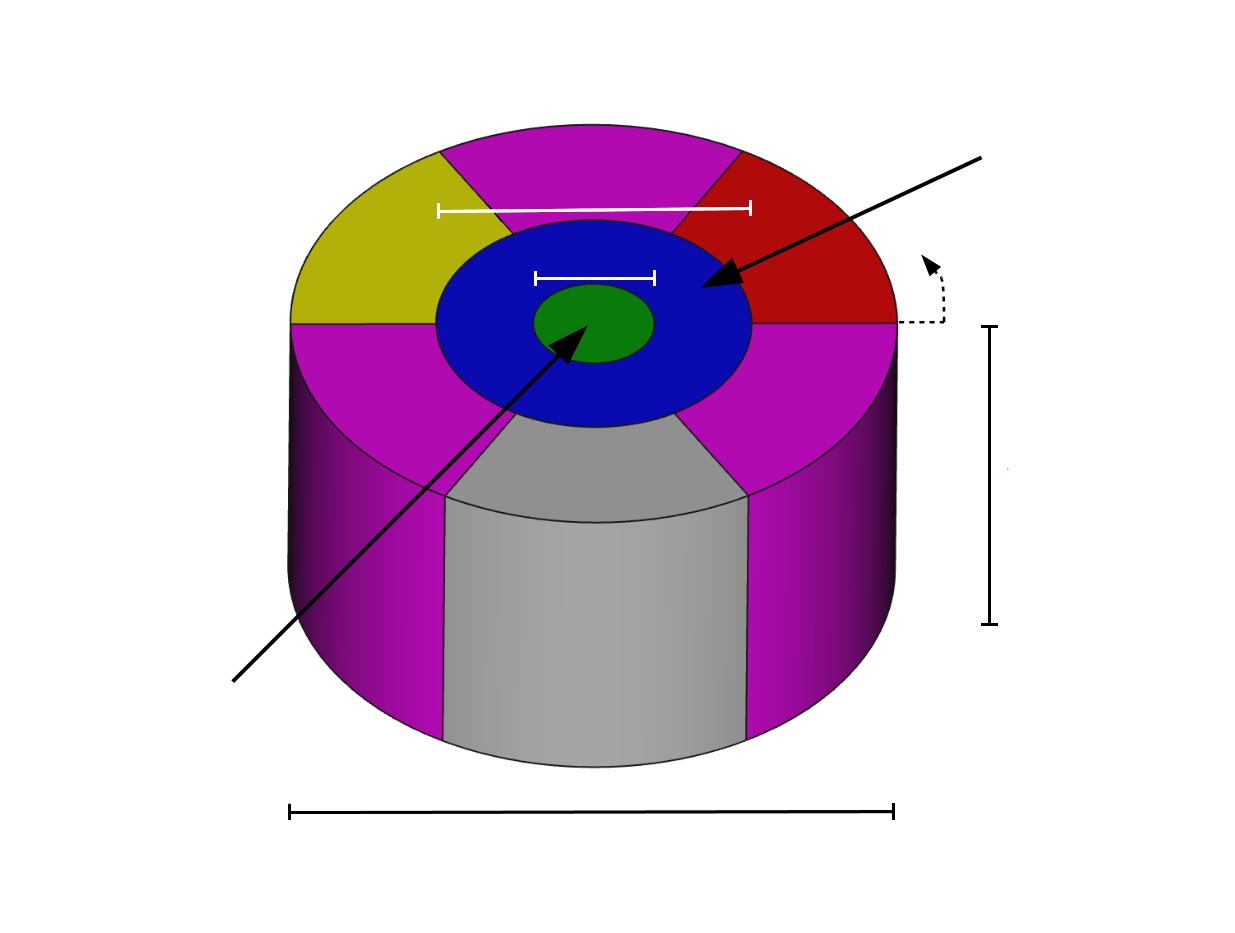}
    \put(41.8,54.6) {\small\color{white}15\,mm}
    \put(41.8,60) {\small\color{white}39\,mm}
    \put(44,4) {\small75\,mm}
    \put(82,37) {\small40\,mm}
    \put(0,15) {\shortstack{Core\\ n++ contact}}
    \put(65,65) {Passivation Area}
    \put(78,52) {$\phi$ = 0$^{\circ}$}
    \end{overpic}}
     \subfloat[Bottom  view.]{\begin{overpic}[width = 0.48 \textwidth,, tics=10]
     {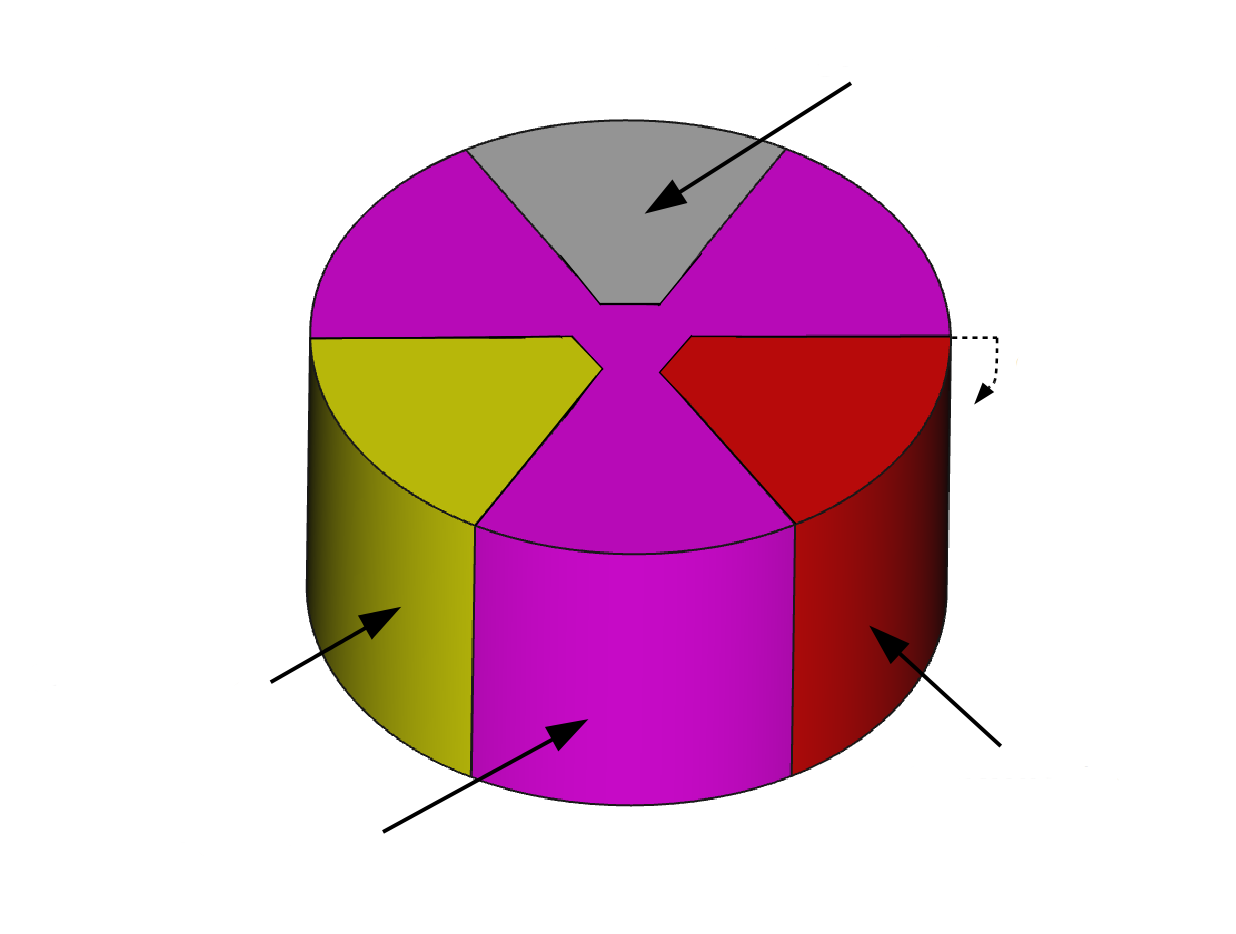}
     \put(-5, 19.0) {Segment 2}
     \put(4, 6.0) {Segment 4}
      \put(74.0, 10) {Segment 1}
      \put(58.0, 71) {Segment 3}
      \put(82,44) {$\phi$ = 0$^{\circ}$}
     \end{overpic}}
     \caption{Schematic of the n-type segmented BEGe detector used to demonstrate the
              capabilities of {\it SolidStateDetectors.jl} throughout the paper.}
     \label{fig:SegBEGePic}
\end{figure}
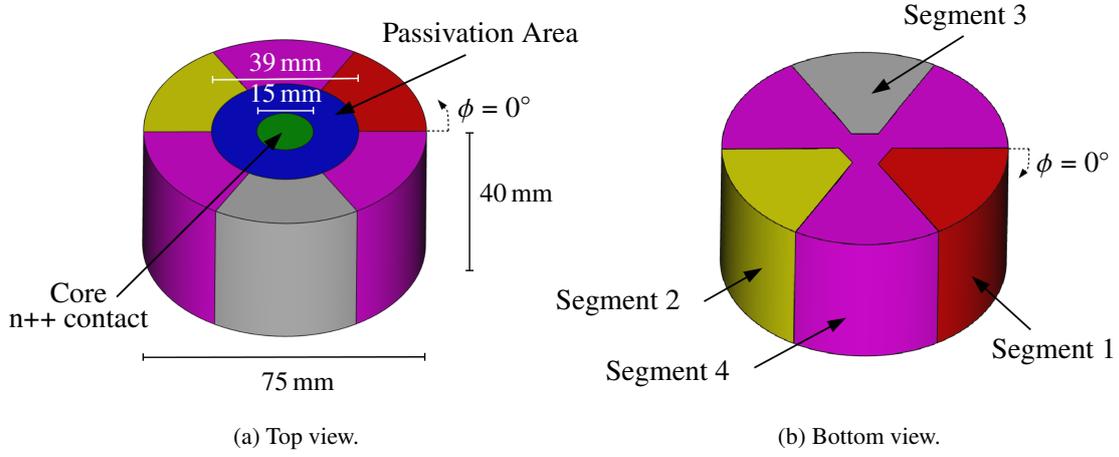

Figure~\ref{fig:SegBEGePic} depicts the n-type segmented BEGe detector~\cite{Abt:2018yvd} which is used as an example throughout this paper. 
This detector features a point contact (core) at the top end-plate.
The core electrode is surrounded by a ring covered with a passivation layer. 
It has a rather unusual segmentation, which was chosen to study the signal development in BEGe detectors and makes it particularly useful to verify a simulation package. 
This segmentation is four-fold: Three equal segments (1=red, 2=yellow, 3=grey) cover about one sixth of the surface each. The remaining surface is covered by one large segment (4=magenta), which is closed at the bottom.
The center of the bottom plate is the origin of a cylindrical coordinate system with the $z$-axis pointing towards the core contact. The left edge of segment~1, looking from the side, defines $\phi = 0^{\circ}$.

\begin{figure}
    \centering
    \begin{overpic}[width=\textwidth, , tics = 10]
       {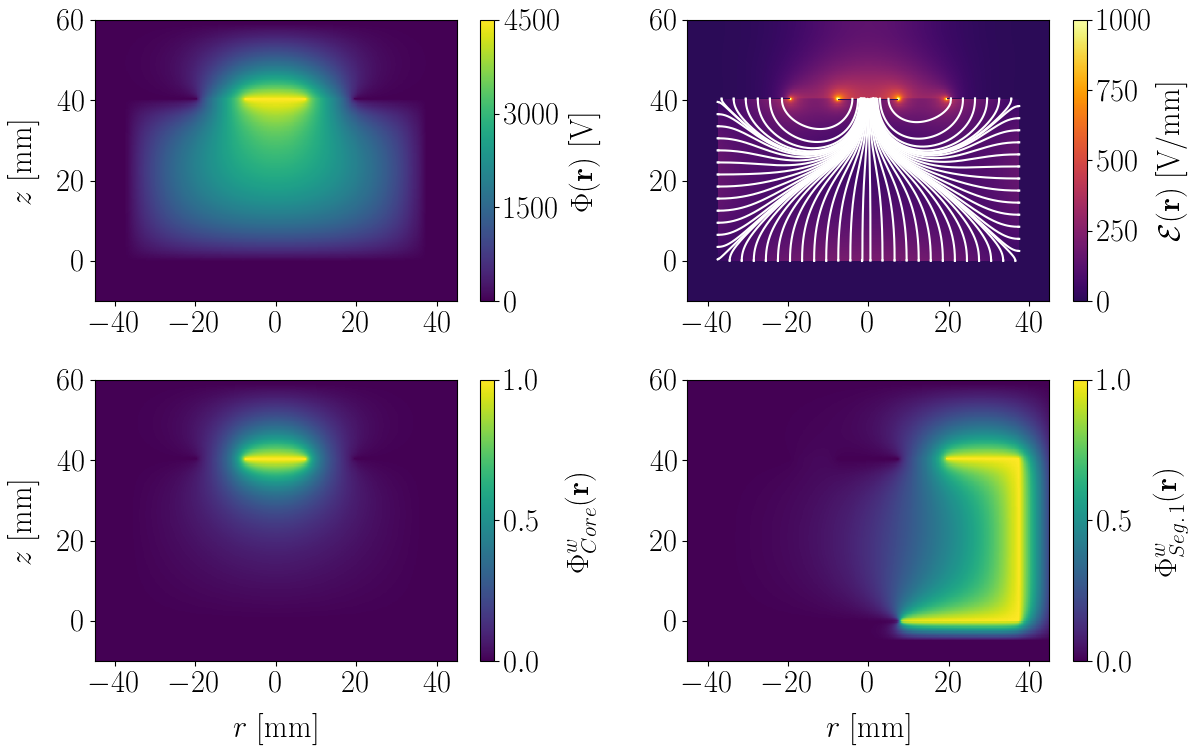}
        \put(1,40)  {a)}
        \put(1,10)  {c)}
        \put(52,40)  {b)}
        \put(52,10)  {d)}
    \end{overpic}
    \vspace{-0.6cm}
    \caption{a) Electric potential, 
    b) electric field strength and field lines,
    c) weighting potential of the core and d) weighting potential of segment~1 
    of the segmented n-type BEGe detector depicted in Fig.~\ref{fig:SegBEGePic}a) at $\phi = 10^{\circ}$ as calculated with {\it SolidStateDetectors.jl}.
    } 
    \label{fig:potentials}

\end{figure}
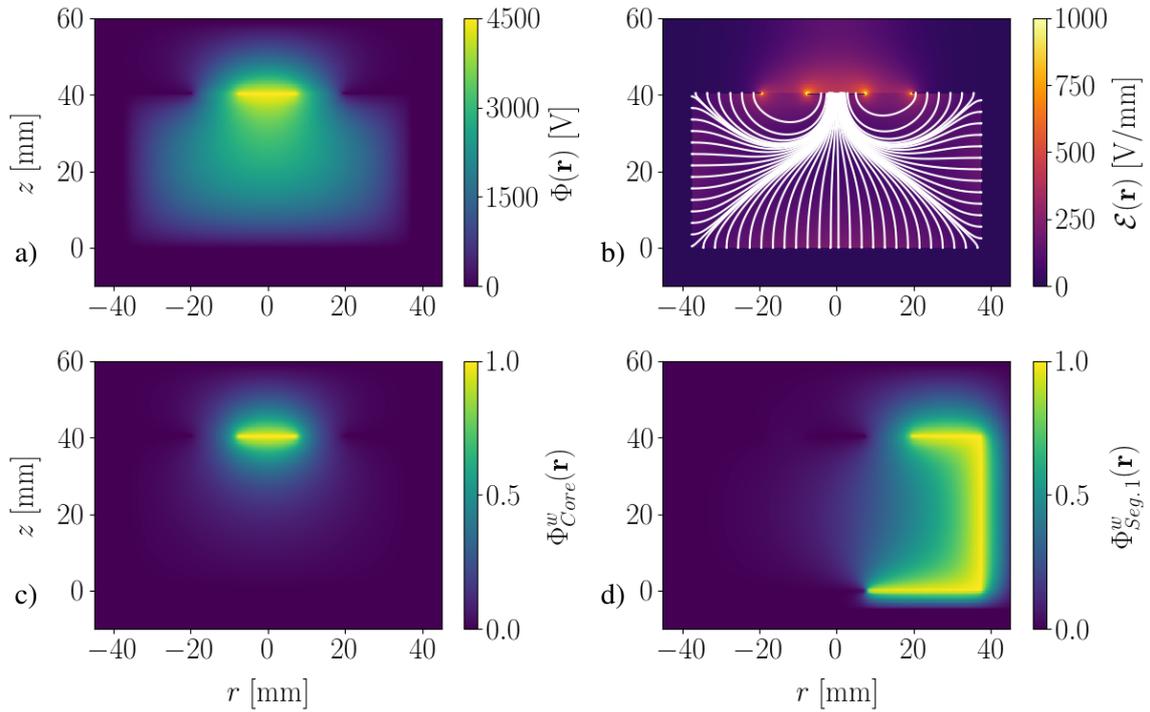

The electric potential and the electric field of the detector as simulated with {\it SolidStateDetectors.jl} are shown in Figs.~\ref{fig:potentials}a) and~b) for the situation that the detector is surrounded by vacuum. 
The high electric field underneath the core contact is typical for BEGe detectors.
Figures~\ref{fig:potentials}c) and~d) show the weighting potentials
for the core and segment~1 at $\phi=10^\circ$.

An infinitely long true-coaxial geometry was implemented 
to validate the numerical field calculation by comparing the result obtained with
{\it SolidStateDetectors.jl} to the analytical solution available for this configuration.
The inner (outer) radius of the germanium detector was set to 5\,mm (35\,mm) and the charge density $\rho(\mathbf{r})$ was defined  as proportional to $r^2$ inside the crystal and zero elsewhere. The potential on the inner (outer) mantle was set to 0\,V (10\,V).
The numerical calculation was performed using both cylindrical and Cartesian coordinates with boundary conditions reflecting in $z$ and periodic in $\phi$.
The symmetries of the detector were used. 
After the default of three refinements, the mean distance between grid points became ($r$: 0.47\,mm, $\phi$: symmetric, $z$: symmetric) for cylindrical coordinates and
($x$: 0.22\,mm, $y$: 0.22\,mm, $z$: symmetric) for Cartesian coordinates, while the
number of grid points became ($r$: 64) and ($x$: 322, $y$: 322). Such a detector would
normally be treated in cylindrical coordinates.
The segmented BEGe detector, see Fig~\ref{fig:SegBEGePic}, was simulated in cylindrical coordinates and after three refinements the number of grid points became 14\,410\,760 ($r$: 259, $\phi$: 260, $z$: 214).

\begin{figure}[htb]
    \centering
    \includegraphics[width=0.95\textwidth]{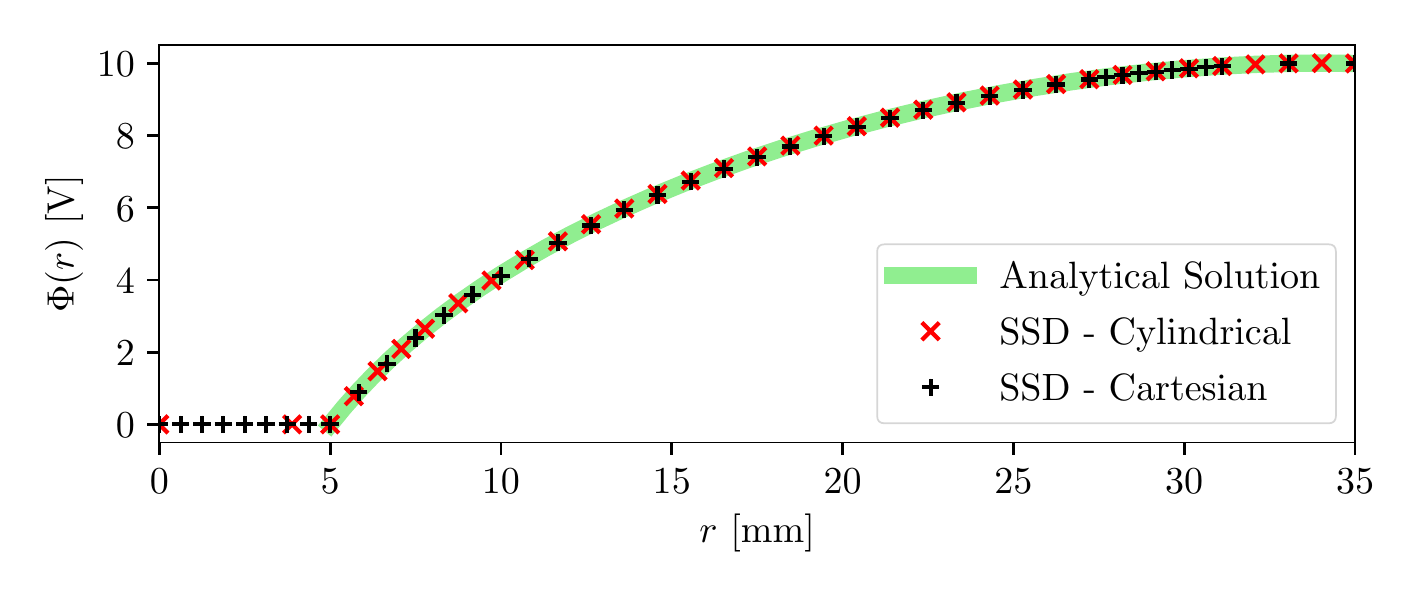}
    \caption{The electric potential in an infinitely long true-coaxial germanium detector as a function of the radius as calculated in cylindrical and Cartesian coordinates with {\it SolidStateDetectors.jl}, SSD, compared to the analytical solution.} 
    \label{fig:validation}
\end{figure}

A comparison of numerically and analytically calculated values of the electric potential is presented in Fig.~\ref{fig:validation}. The root mean square,~$RMS$, difference between the numerical and analytical solutions is very small for the calculation performed in Cartesian coordinates,~$RMS = 0.025$\,V, and almost zero for cylindrical coordinates,~$RMS = 2.1\cdot10^{-6}$\,V.
These results were obtained with the automated termination of iterations and grid refinements.
In both cases, the $RMS$ could be further reduced if iterations and grid refinements beyond the defaults were initiated. 
However, the precision obtained with the defaults would be sufficient for any real-life detector.

The capacitance, $C$, of a detector can be calculated as $C = 2W/V_B^2$, where $V_B$ is the applied bias voltage and $W$ is the energy stored in the generated electric field:
\begin{equation}
    W = \dfrac{1}{2}\epsilon_{0} \iiint\limits_{\mathbb{R}^3} \epsilon_{r}(\mathbf{r}) |\bm{\mathcal{E}}(\mathbf{r})|^2 \mathrm{d}\mathbf{r}.
\end{equation}
The integral becomes a sum over all grid points of the world. 

The procedure is verified by comparing numerically to analytically calculated capacitance
values for the infinitely long true-coaxial detector and, in addition, for an infinitely long and wide parallel plate detector.
The ratios of the numerical (cyl = cylindrical coordinates, car = Cartesian coordinates) and analytical (true) values are
    $C_{\mathrm{cyl}}^{\mathrm{coax}} / C_{\mathrm{true}}^{\mathrm{coax}} = 0.9997$,
    $C_{\mathrm{car}}^{\mathrm{coax}} / C_{\mathrm{true}}^{\mathrm{coax}} = 1.01$,
    $C_{\mathrm{car}}^{\mathrm{plate}} / C_{\mathrm{true}}^{\mathrm{plate}} = 1.006$.
These ratios asymptotically approach unity if the number of iterations and grid refinements is increased beyond the default values.
The comparisons described here are part of the automatic tests, which are performed every time the source code of {\it SolidStateDetectors.jl} is changed.

\begin{figure}[htp]
    \centering
    \begin{overpic}[width=0.98\textwidth, , tics = 10]
    {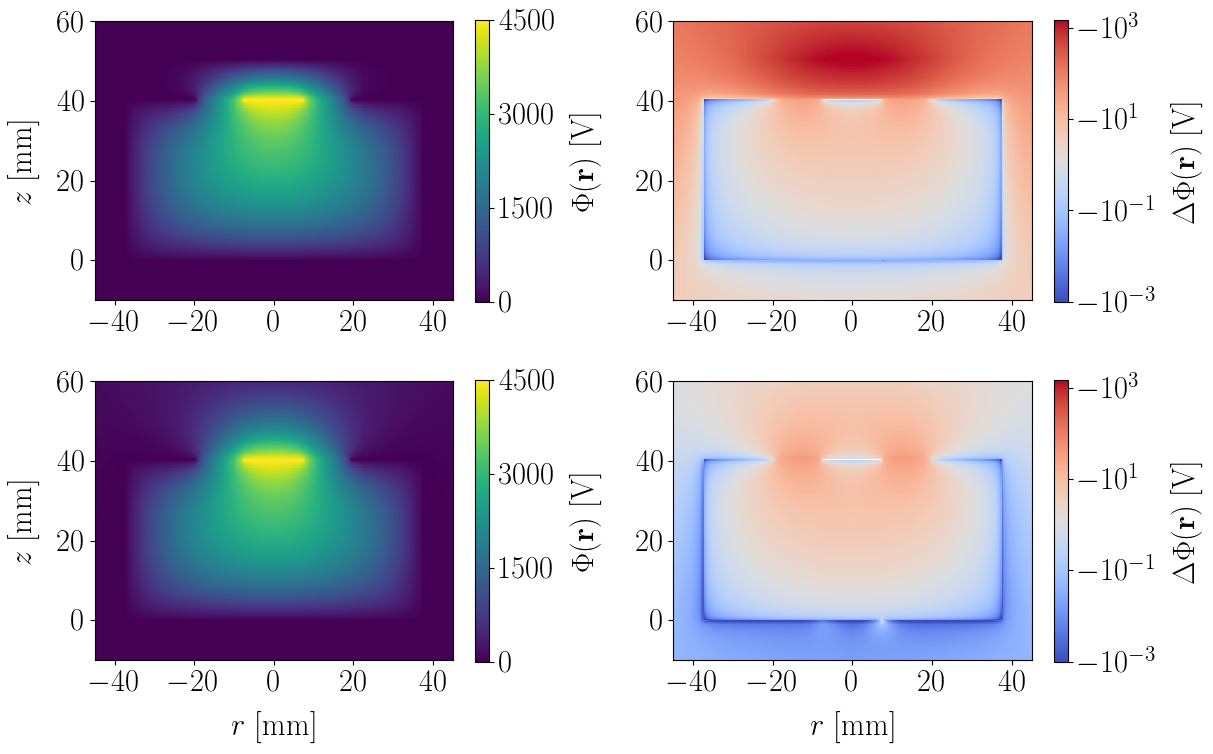}
    \put(2,38) {a)}
    \put(50,38) {b)}
    \put(2,8) {c)}
    \put(50,8) {d)}
    \put(10,40) {\color{white} 2mm Copper shell}
    \put(10,10) {\color{white} Surrounded by LAr}
    
    \end{overpic}
    \caption{a) and c) Electric potential inside the world volume simulated for the n-type segmented BEGe detector shown in Fig.~\ref{fig:SegBEGePic}a) as calculated for
    a) the scenario of a 2\,mm thick copper shell surrounding the top and the side of the detector at a distance of 10\,mm from the surface and c) the detector being submerged
    in liquid argon; b) and d) 
    respective differences to the values obtained for the detector being surrounded by vacuum: $\Delta\Phi(\mathbf{r}) = \Phi_{\mathrm{Cu/LAr}}(\mathbf{r})-\Phi_{\mathrm{Vacuum}}(\mathbf{r})$. 
    }
    \label{fig:surrounding}
\end{figure}
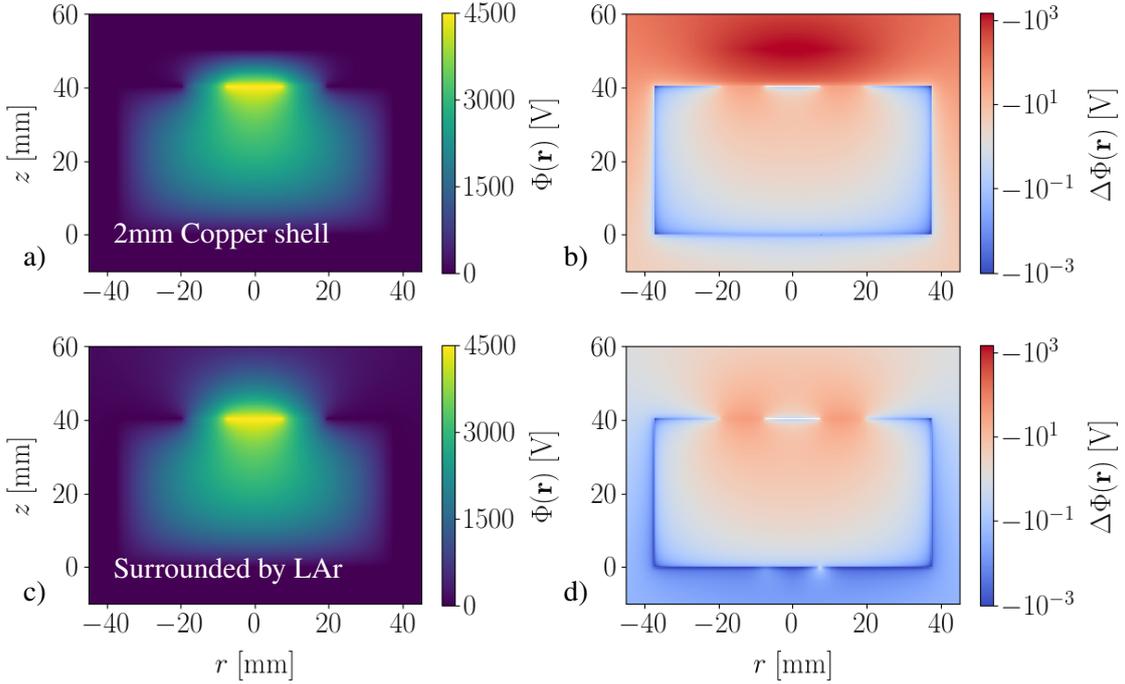

In {\it SolidStateDetectors.jl}, the relative permittivity $\epsilon_{r}(\mathbf{r})$ is a position dependent parameter.
This allows the user to implement the surroundings of the detector, e.g.\ infrared shields or support structures in close proximity to the crystal. 
While HPGe detectors are typically operated in vacuum cryostats, collaborations like 
{\sc Gerda}~\cite{Ackermann:2012xja} or {\sc LEGEND} submerge their detectors directly in liquid argon.
Figure~\ref{fig:surrounding} demonstrates the importance of taking the surroundings of a detector into account.
Especially, volumes close to the passivated surfaces exhibit non-negligible differences in the electric potential.
The two cases investigated here, a copper shell and submersion in liquid argon, result in a similar reduction of the potential inside the detector underneath the passivated ring. However, the effect is stronger for the detector submerged in argon. 
The effect outside the detector is even stronger than inside, especially for the copper shell. However, the effect is particularly important for the case of submersion in liquid argon, where free radioactive ions can be attracted to the detector.  

Alpha and beta decays of contaminants on the passivated surfaces of germanium detectors are known to be critical sources of background for rare event searches. Due to the special field conditions underneath these surfaces such events can look like $0\nu\beta\beta$ events. 
Thus, the knowledge on possible changes in the electric field and in the behavior of the detectors due to the surrounding medium, surrounding support structures and cables is important for the next generation of such experiments, which rely on a substantial reduction of the background level compared to what has been achieved so far.

\section{Charge drift and induced signals}
\label{sec:drift}

The drift velocity vectors are calculated separately for electrons and holes for each grid point using the electric field and the respective electron or hole drift velocity model.
The default drift velocity models implemented in {\it SolidStateDetectors.jl} are based on the AGATA Detector Library~\cite{Bruyneel:2006764, Bruyneel:2016zih} and are described in some detail in Appendix~\ref{app:drift}.

For each energy deposition, the drift paths are calculated for point-like electron and hole clouds, using drift velocities, $\mathbf{v}_{e/h}(\mathbf{r}_{e/h})$, as interpolated linearly from the neighboring grid points.
The drift vector for each step is calculated as $\Delta\mathbf{r}_{e/h} = \mathbf{v}_{e/h}(\mathbf{r}_{e/h}) \, \Delta t$, where the step length, $\Delta t$, can be chosen by the user and is usually in the range of 1--10\,ns.
The drift of a charge ends when it reaches an electrode or a maximum number of steps have occurred.
If $|\Delta\mathbf{r}_{e/h}|$ is below a certain limit the drift can also end inside the detector. 
For steps leading beyond the detector volume, the intersection between the drift vector and the detector edge is determined.
If the crossing point does not belong to an electrode, the drift vector for this step is reduced to its component parallel to the surface. 
The charge drift ends when the crossing point belongs to an electrode.

Charges can get trapped inside the bulk of a germanium detector or, more commonly, underneath a passivated surface. The former is predicted by {\it SolidStateDetectors.jl} for volumes where the electric field is close to zero. The latter can, in general, not be predicted. There are several known effects like a surface charge-up or a surface channel, but there is no complete model for them. The user can, however, define virtual volumes, in which the drift model can be modified to simulate such effects. This is usually done after some effects have shown up in the data, which need to be understood. In many cases, a reduction of drift speed along the surface provides a first insight into what is happening. 

\begin{figure}
    \centering
    \begin{overpic}[width=0.98\textwidth,,tics = 10]
    {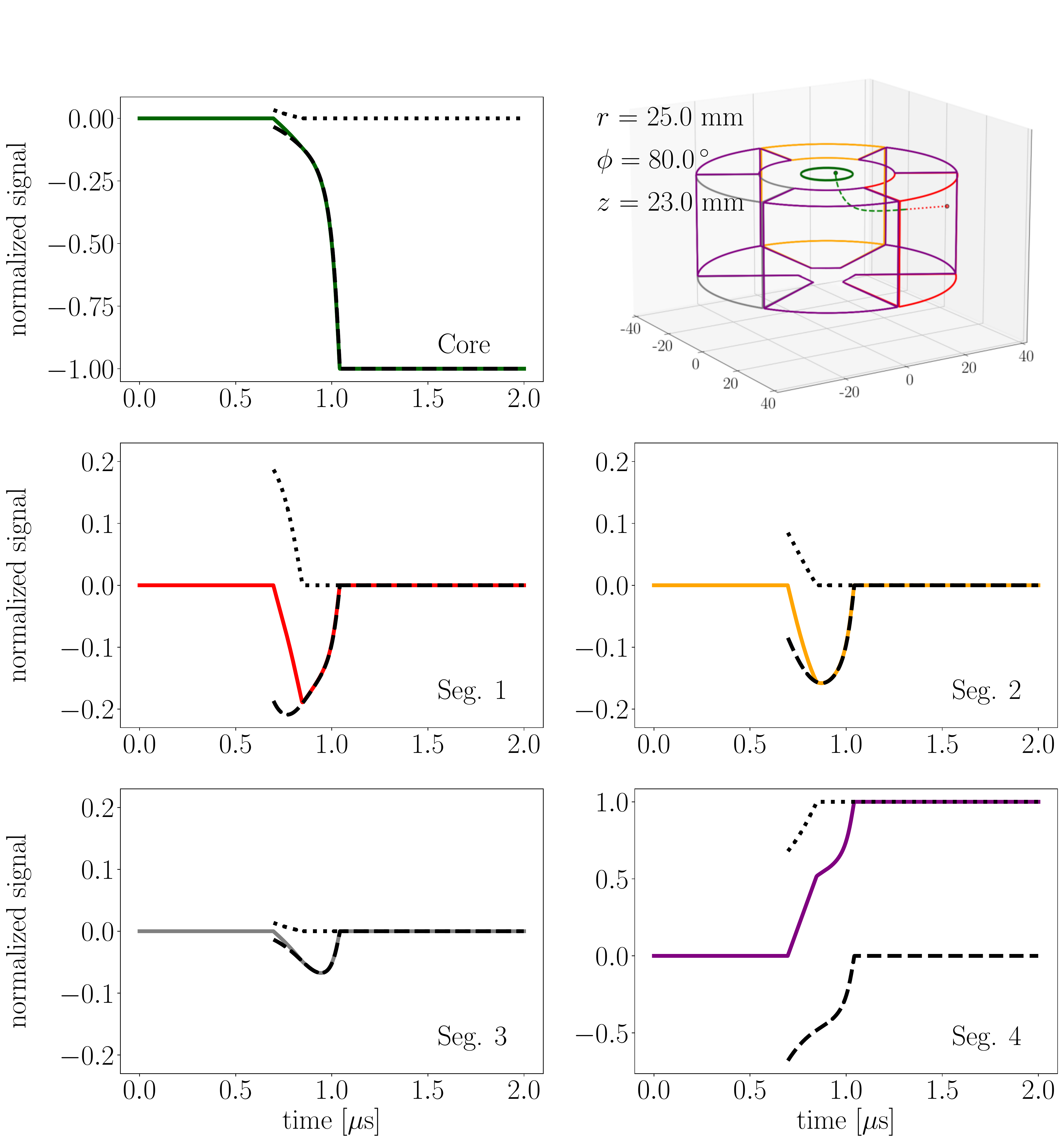}
    \put (76,63) {\small\rotatebox{11}{$y$\,[mm]}}
    \put (55,68) {\small\rotatebox{-35}{$x$\,[mm]}}
    \end{overpic}
    \caption{Drift trajectories of electrons (dashed line) and holes (dotted line) for an energy deposition in the bulk of the n-type segmented BEGe detector at  $r=25.0$\,mm, $z=23.0$\,mm, and $\phi=80.0^{\circ}$ as well as the time dependent signals (charges) induced in the core and the segment electrodes. The contributions from the electrons (holes) are shown as dashed (dotted) lines.
  }
    \label{fig:simevent}
\end{figure}

The electric signals are the sum of the charges induced in the electrodes of the detector by all drifting electrons and holes.
The Shockley-Ramo theorem~\cite{Schockley1938,Ramo1939,SchockleyRamo2001} is used to calculate the time development of the induced charge, $Q_i(t)$, in each electrode $i$ from $\mathbf{r}_{e/h}(t)$ and $\Phi^w_i(\mathbf{r})$:
\begin{equation}
    Q_i(t) = Q_0 \left[ \Phi^w_i(\mathbf{r}_h(t)) - \Phi^w_i(\mathbf{r}_e(t)) \right]\mathrm{,}
\end{equation}
where $Q_0$ is the absolute electric charge in the electron and hole clouds after separation of the charge-carrier pairs.

Figure~\ref{fig:simevent} shows the simulated drift trajectories and the signals induced in all electrodes for an energy deposition in the bulk of the n-type segmented BEGe detector. 
The separate contributions from electrons and holes show that the signal of the core electrode is dominated by the electron drift in this type of detector due to the strong gradient of the weighting potential around the small core electrode, see Fig.~\ref{fig:potentials}c).

A positive collected charge is observed for segment~4, which is the collecting segment for this event. 
The other segments show mirror pulses that return to zero after the drift is completed.
The amplitudes of the mirror pulses in segments~1 and~2 indicate that the position of the energy deposition is closer to segment~1 than segment~2. 
The possibility to obtain 3D position information from fits to the events in simulated pulse-shape libraries including all mirror pulses will be investigated in the future.

The parameters which are used to calculate the drift velocities, see Eq.\,(\ref{eq:vl}) in Appendix~\ref{app:drift},
are temperature dependent. The default values used in {\it SolidStateDetectors.jl}
were determined for the reference temperature of 78\,K~\cite{Bruyneel:2006764, Bruyneel:2016zih}. However, germanium detectors are often operated at other temperatures.
The effect of the temperature on the drift velocities can be taken into account
by applying user defined
model-dependent functions as described in Appendix~\ref{app:temp}.

\section{Comparison of simulation to data}
\label{sec:validation}

Data from the n-type segmented BEGe detector, see Fig.~\ref{fig:SegBEGePic}, were used to validate the simulation. 
The detector was mounted with the core electrode on top in an electrically cooled aluminum vacuum-cryostat with a continuously monitored temperature control system~\cite{Hagemann-mt}.
The detector was irradiated in three measurement campaigns with $^{133}$Ba, $^{137}$Cs and $^{228}$Th.

The signals from the core and segment electrodes were amplified with charge-sensitive preamplifiers and digitized with a sampling rate of 250~MHz. The core signal was inverted.
The recorded pulses were baseline subtracted and corrected for their preamplifier channel-specific decay.
Signal amplitudes were derived using a fixed-size window filter. 
Linear cross-talk between the core and the segments as well as between segments was corrected for in an automated energy-calibration procedure using calibration parameters obtained from single-segment events~\cite{Schuster-mt,Abt:2018yvd,Hauertmann-mt}.
The calibration was performed for each data set individually.
The data were not corrected for differential cross-talk. This kind of cross talk does not affect the measured energy but changes the shape of the affected pulses. A model based on the first derivative of the charge pulses, the currents, was developed~\cite{Hagemann-mt} and applied to the simulation before a comparison to data\footnote{The complete cross-talk model is that the vector of measured channel pulses, $p^m(t)$, at each time, $t$, is a linear combinations of all true pulses, $p^t(t)$ and their first derivatives, $p^m(t) = CL * p^t(t) + CD * dp^t(t)/dt$, where $CL$ is the linear and $CD$ the differential cross-talk matrix.}. 

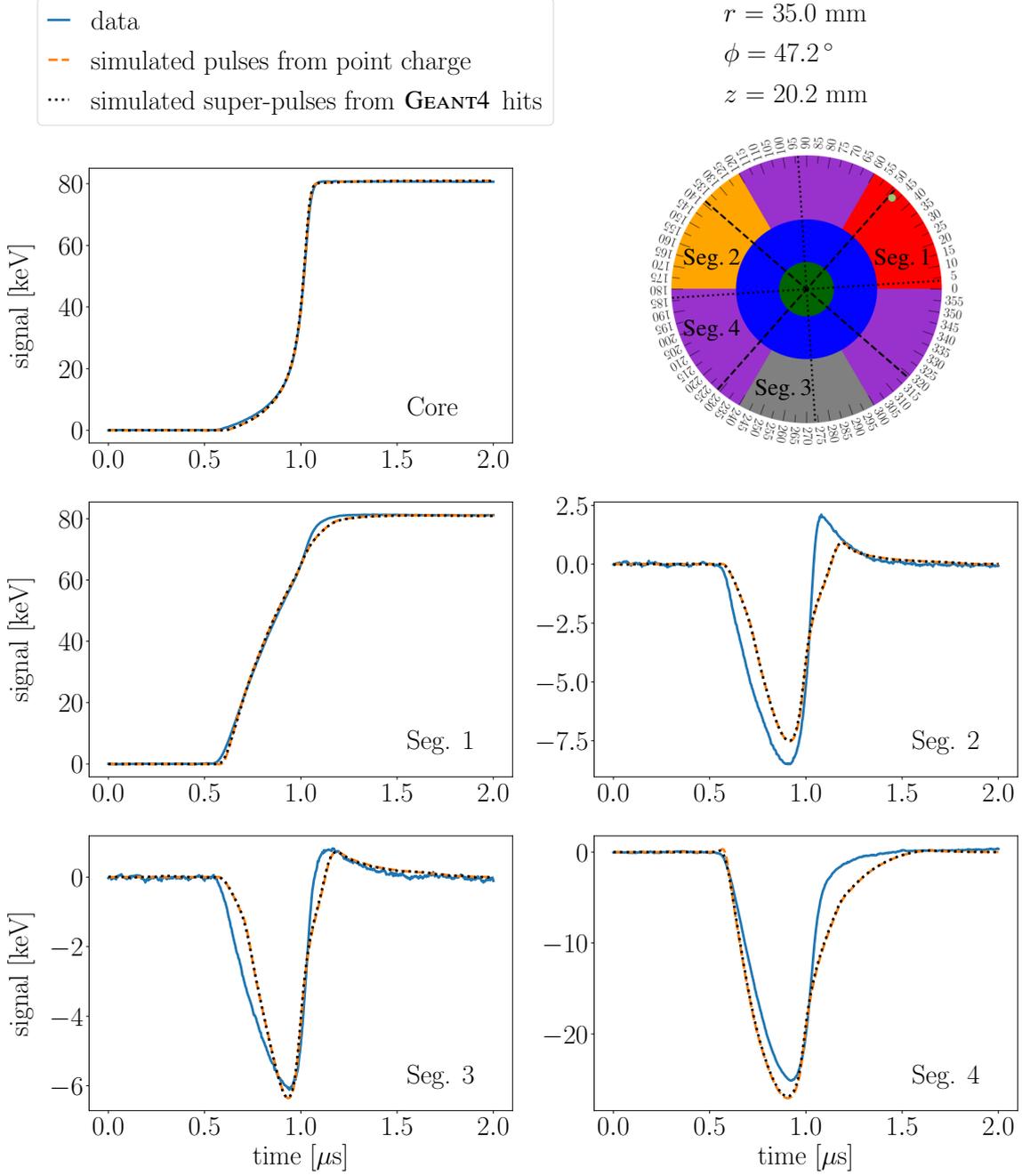
\begin{figure}
    \centering
    \begin{overpic}[width=\textwidth,, tics = 10]
    {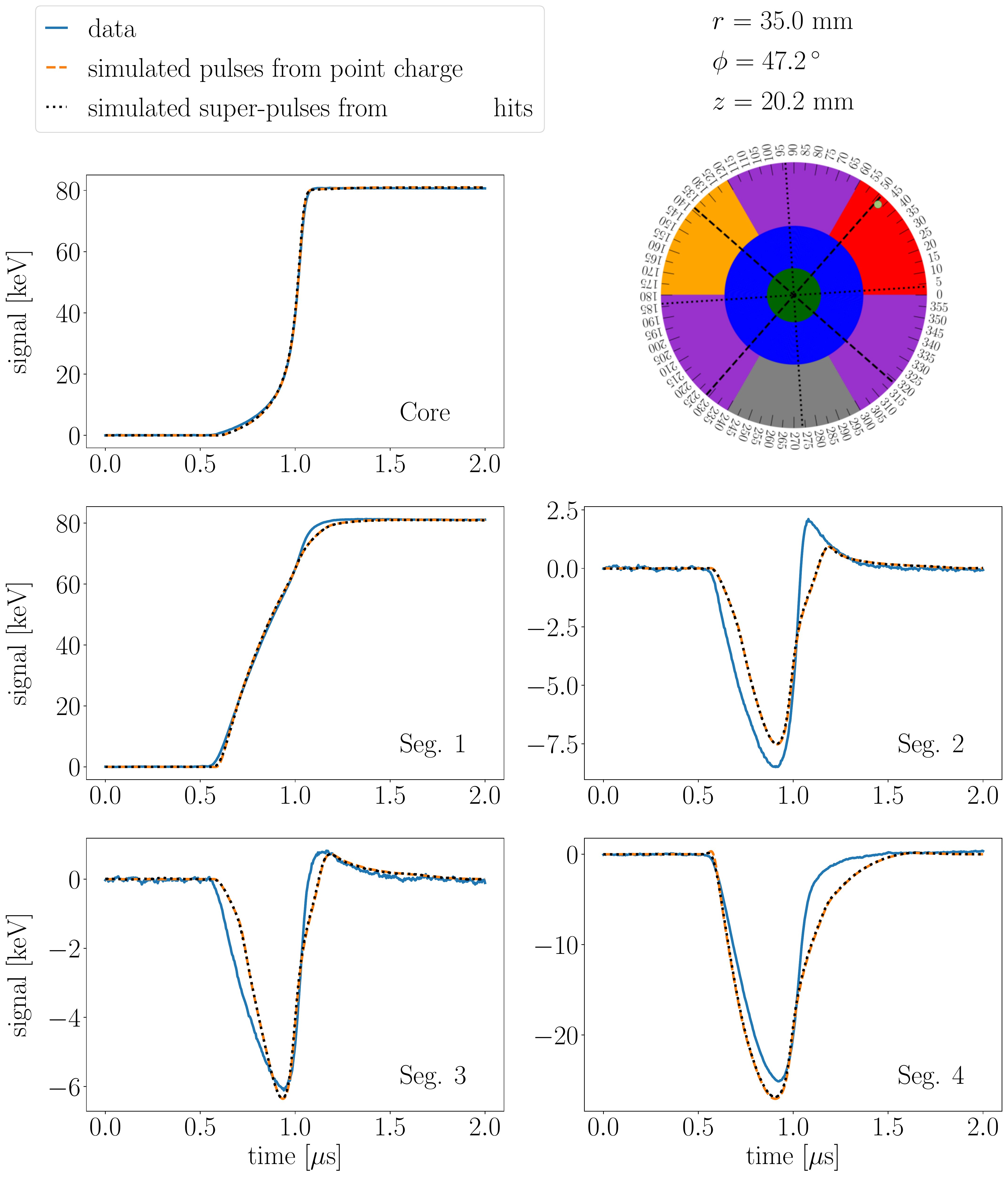}
    \put(73.1,77) {\small Seg.\,1}
    \put(57.1,77) {\small Seg.\,2}
    \put(57.1,71) {\small Seg.\,4}
    \put(63.1,66) {\small Seg.\,3}
    \put(33.8,90.2) {{\scshape Geant4}}
    \end{overpic}
    \caption{Measured and simulated super-pulses for interactions of 81\,keV photons 
    from a collimated $^{133}$Ba source, placed on the side of the detector irradiating
    segment~1 at $z=20.2$\,mm and $\phi=47.2^{\circ}$. The simulation is shown for charge injection at the average depth of 2.5\,mm and a full {\sc Geant4} simulation. The top right insert provides the segment numbering and the $\phi$ position of the interaction point on the $\langle100\rangle$ axis. Differential cross-talk was applied to the simulation~\cite{Hagemann-mt}.
    The scales of the y axes for the mirror pulses were chosen such that remaining small differences between simulation and data become visible.}
    \label{fig:pulseBa}
\end{figure}

Each experimental setup, including the respective collimator and source encapsulation, were implemented in {\sc Geant4}. 
The hits in the detector as simulated with {\sc Geant4} were clustered and the induced charges were drifted using {\it SolidStateDetectors.jl}.
The simulated pulses for the core and all segments were convolved with the response function of the respective preamplifier\footnote{Charges always drift to the core or a segment electrode; the segment boundaries do not cause any deficit in charge collection.}.
The response functions were measured by injecting test pulses with a rise-time of a few nanoseconds as input to the preamplifiers and recording the resulting pulses.
The decay of the preamplifier pulses was included in the response function and was varied within uncertainties to reproduce the variations observed in the data.
Furthermore, the observed baseline noise was added to the simulated pulses.
The resulting pulses were used for comparisons to calibrated and cross-talk corrected data.

\subsection{Surface events}

Photons from the 81\,keV line of $^{133}$Ba were used to study events close to the detector surface. The data were collected by placing a collimated $^{133}$Ba source on the side of the detector, which was kept at the reference temperature of 78\,K.
The pulses from single-segment events in the $(81\pm2)$\,keV energy interval were averaged to form so-called super-pulses in order to reduce the effect of noise and to average out the spatial distribution of events due to the size of the beam spot.
Figure~\ref{fig:pulseBa} shows a comparison of simulated to measured super-pulses for
a measurement where the middle of the mantle, $z=20.2$\,mm, was irradiated at $\phi=47.2^{\circ}$.
The super-pulses from simulation describe the data reasonably well.
The core pulse is very well simulated. The simulated pulse of the collecting segment shows a slow-down at the end which is not observed in data and which is reflected in the mirror pulses as simulated for the non-collecting segments.  
The average penetration depth of 81\,keV photons in germanium is approximately 2.5\,mm such that the energy was deposited close to the surface and the holes were collected quickly. The electrons drift inwards along the $\langle100\rangle$ axis and upwards along the $\langle001\rangle$ axis. As the drift was dominated by electrons, the results shown in
Fig.~\ref{fig:pulseBa} validate the simulation procedure and the drift velocity model for electrons on the relevant axes at a temperature of 78\,K.

The simulation is shown for the case where charges were injected at the nominal 
$(z,\phi)$ position at a depth of 2.5\,mm and the case 
where the charges induced by the individual hits as provided by {\sc Geant4} were drifted separately. 
The isotope $^{133}$Ba also features four gamma lines at higher energies (276, 303, 356 and 384\,keV). These gammas penetrate deeper into the detector and can create background events with a deposited energy of around 81\,keV through Compton scattering. The {\sc Geant4} simulation included these gammas. As for the data, such events were included in the super-pulse formation.
Nevertheless the super-pulses from charge injection and {\sc Geant4} simulation are  basically indistinguishable. This confirms that the signal to background ratio is high enough and that the detailed shapes of the initial charge clouds are effectively averaged out by forming super-pulses for low energies.
The remaining small differences between data and simulation observed for the segments could be due to the lack of knowledge about the exact impurity density distribution in the detector\footnote{The impurity density distribution as provided by the manufacturer was multiplied with 0.9 to adjust the overall pulse length along the $\langle100\rangle$ axis. This is well within the uncertainty on the provided impurity density.  No radial profile was implemented.}. 
They could also be a hint that the parameters of the electron drift model are slightly off.
Such effects will be investigated in the future. 
Differential cross-talk was applied to the simulation and is unlikely to be a major source of remaining differences, see next section.

\subsection{Bulk events}

Data taken with a Compton scanner~\cite{Hagemann-mt} using a
collimated $^{137}$Cs source mounted above the detector were used to compare simulated to 
measured pulses from energy depositions in the bulk of the detector.
The Compton-scattered photons from the 662~keV line were detected by a cadmium zinc telluride (CZT) pixel detector with a position resolution of about $0.5$\,mm
and an energy resolution of about 1\% at the relevant energies.
Events were selected which had coincident energy depositions in the BEGe and CZT detectors with a total energy of $(662\pm5)$\,keV.

Events with two hits in the CZT detector were used to select
energy deposits in specific regions of the BEGe detector within $(z\pm1)$\,mm, using
a technique based on the reconstruction of Compton cones~\cite{Hagemann-mt}.
The $r$ and $\phi$ coordinates were taken from the source position.
Super-pulses were formed by an energy-weighted average of the pulses from the selected events after normalization.
Additional pulses from events with one hit in the CZT detector, which were compatible with the previous super-pulses,
were added to the averaging for the final super-pulses of the chosen volume to further reduce the effect of noise in the data.

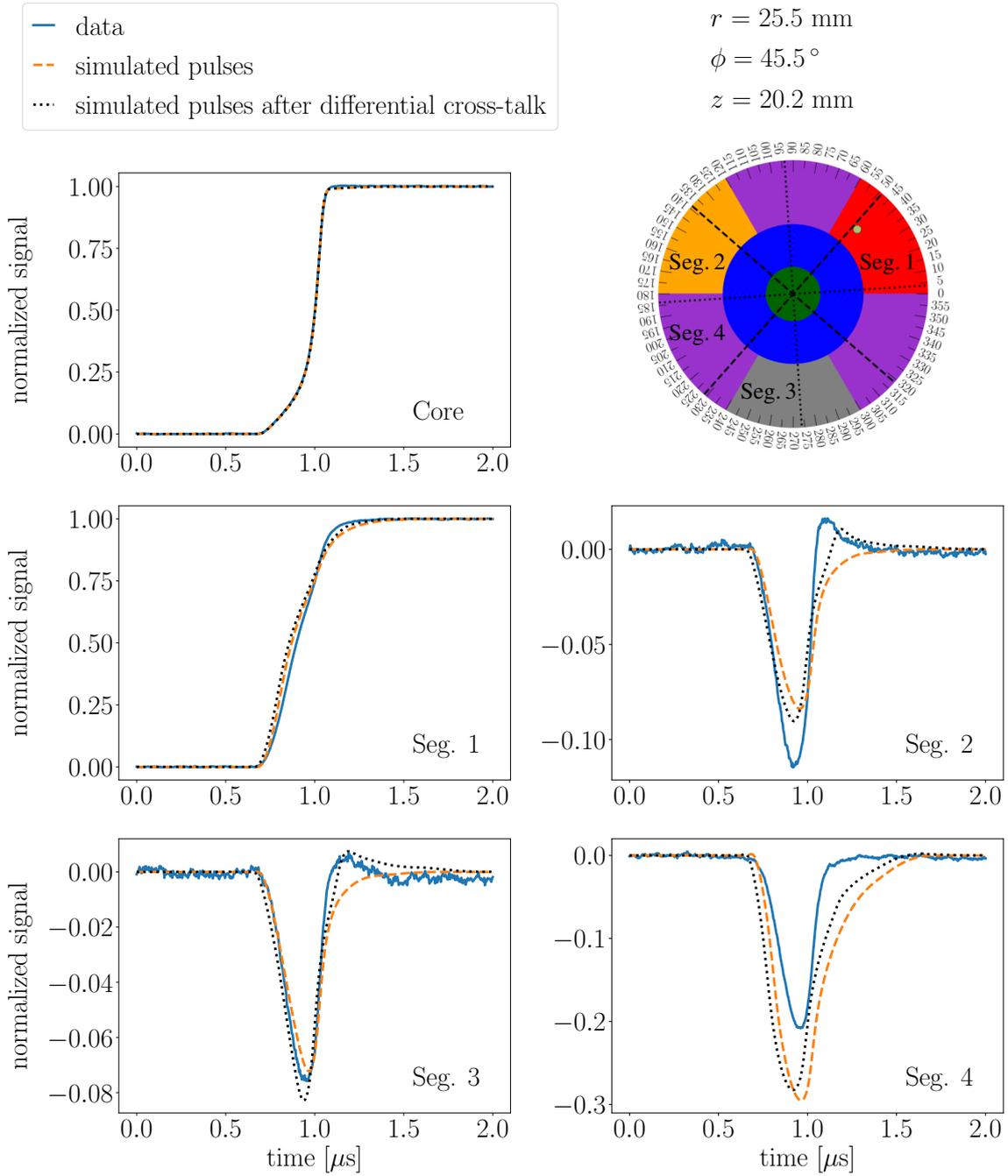
\begin{figure}
    \centering
    \begin{overpic}[width=\textwidth, , tics = 10]
    {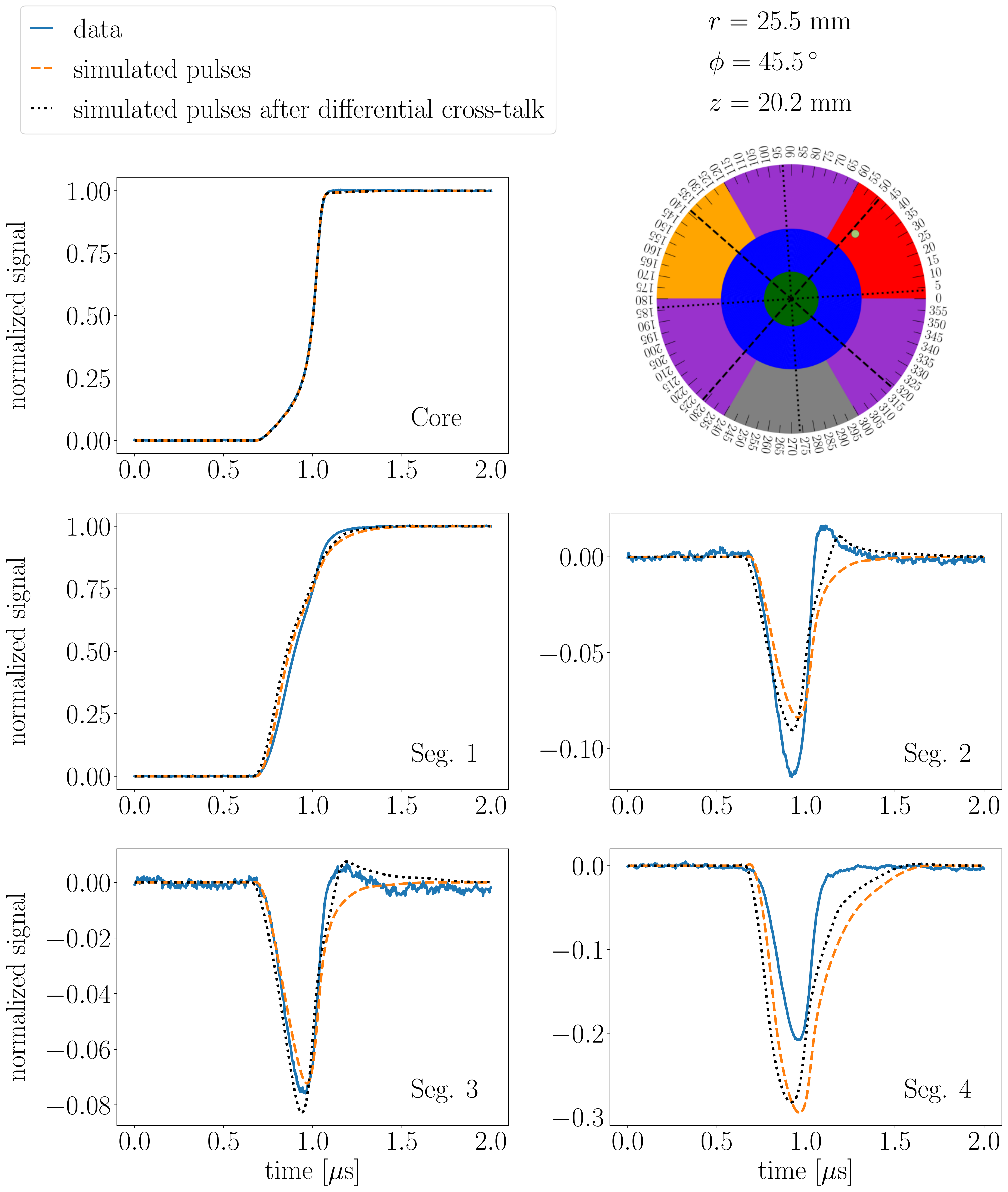}
    \put(72,77) {\small Seg.\,1}
    \put(56,77) {\small Seg.\,2}
    \put(56,71) {\small Seg.\,4}
    \put(62,66) {\small Seg.\,3}
    \end{overpic}
    \caption{Simulated super-pulses compared to super-pulses from Compton reconstructed events in the bulk of the detector ($r=25.5$~mm, $\phi=45.5^{\circ}$, $z=20.2$~mm). The data is corrected for linear cross-talk. 
    The simulation is shown with and without the application of differential cross-talk.}
    \label{fig:pulseCs}
\end{figure}

Figure~\ref{fig:pulseCs} shows the simulated super-pulses for an example location at $r=25.5$\,mm compared to the data.
The simulation agrees reasonably well with the data.
This validates the simulation in the bulk of the detector, 
where the hole drift also plays an important role in the pulse formation.
The data is corrected for linear cross-talk. The simulation is shown with and without the application of differential cross-talk~\cite{Hagemann-mt}. Differential cross-talk is significant here because the time-structure of the pulses for the core and the collecting segment are sufficiently different. 
It mostly affects the segment pulses while the core pulse is basically not changed.
The application of differential cross-talk improves the description of the data by the simulation, especially the description of the mirror pulses.
The residual differences between the simulated and measured pulses of the collecting segment might be due to the lack of knowledge on the exact impurity distribution, which is an input to the simulation. 
They could also be an indication that the input parameters to the hole and the electron drift models need to be adjusted. The relative mobilities  of holes and electrons are not
very well known for germanium.
Another effect to be investigated is diffusion and self-repulsion of the charge cloud. This will be implemented in {\it SolidStateDetectors.jl} in the near future.

Overall, the description of the data by the simulation is quite good. A quantitative statement can be obtained using pulse-shape analysis. 

\subsection{Pulse-shape analysis}

One of the most common pulse-shape analysis techniques for point-contact detectors is based on the so-called $A/E$ parameter, where $A$ is the maximum of the differentiated core pulse (i.e.\ current) and $E$ is the energy of the event.
The performance of the simulation was tested using data, for which an uncollimated $^{228}$Th source illuminated the detector from the side. It was placed 20\,mm away from the surface at $z=20.2$\,mm at the center of segment~1.
Core pulses from a measurement lasting about 9~hours were used to measure the $A/E$ distribution in selected energy regions.
The simulated core pulses of events produced with {\sc Geant4} for the experimental setup were smoothed and numerically differentiated exactly like the pulses recorded for data.

\begin{figure}
    \centering
    \begin{overpic}[width=\textwidth, , tics = 10]
    {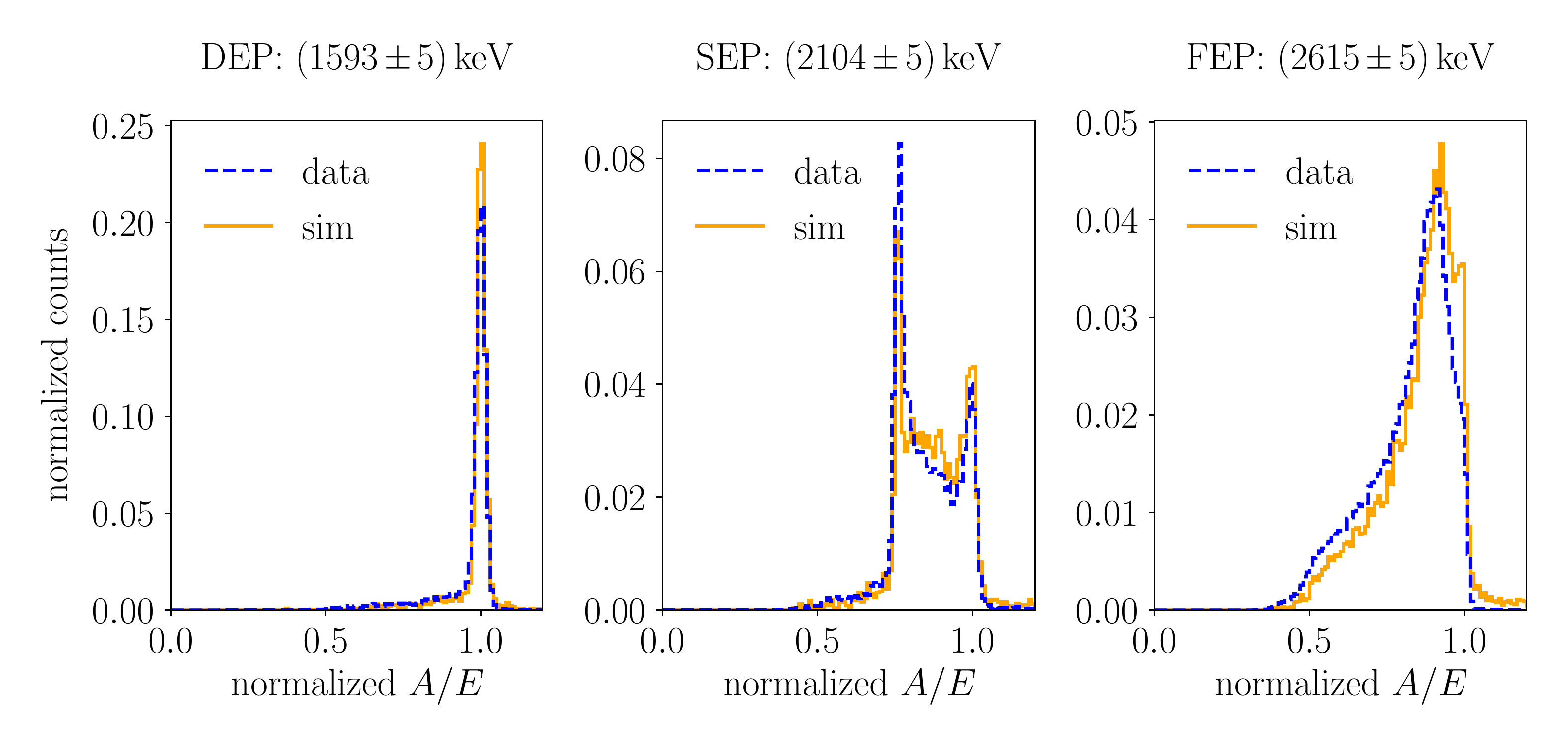}
    \put (6,3)  {a)}
    \put (37.5,3) {b)}
    \put (69,3) {c)} 
    \end{overpic}
    \caption{Simulated $A/E$ distributions compared to data for the a) double escape, b) single escape and c) the full energy peak from 2615\,keV photons originating from $^{208}$Tl decay.}
    \label{fig:aovere}
\end{figure}

In Fig.~\ref{fig:aovere}, a comparison of the simulated $A/E$ distributions to data and is shown for events in the double escape peak (DEP), single escape peak (SEP) and full energy peak (FEP) from 2615\,keV photons from the decay of $^{208}$Tl.
The position of the peak in the $A/E$ distribution derived from the $^{208}$Tl DEP was scaled to 1 for data and simulation.
The scaling factor for the data was only 0.5\% larger than for the simulation.
The simulated distributions describe the data quite well. However, the simulated distribution of $A/E$ for the FEP shows events which have $A/E$ larger than 1. Such events
do not occur in data. They originate from large energy depositions directly underneath the core contact. This will be studied further. 

\begin{figure}
    \centering
    \begin{overpic}[width=\textwidth, , tics = 10]
    {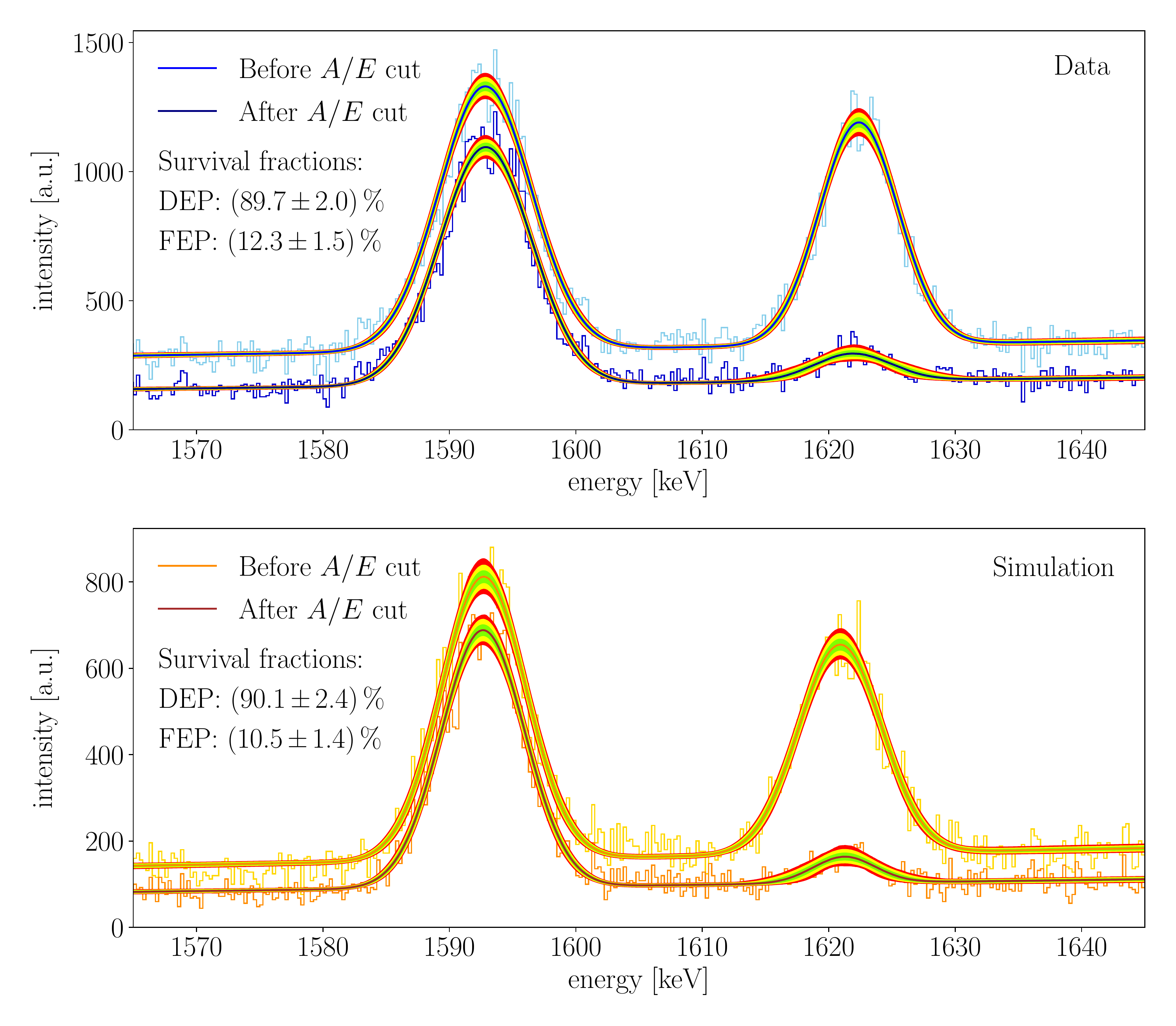}
    \put (3,48) {a)}
    \put (3,5) {b)}
    \end{overpic}
    \caption{Energy spectra from events from $^{228}$Th a) data and b) simulation before and after the application of an $A/E$ cut providing
    a survival rate of $\approx90\%$ for the $^{208}$Tl DEP.
    Shown is the energy range around the $^{208}$Tl DEP and the $^{212}$Bi FEP. 
    Survival rates were determined with fits performed with {\it BAT.jl}~\cite{Schulz:2020ebm}.
    The confidence intervals $1\sigma$, 2$\sigma$ and $3\sigma$ are shown for all fits.}
    \label{fig:psdcut}
\end{figure}

The DEP from $^{208}$Tl was used to determine a cut in 
$A/E$, which separates single-site from multi-site events. 
The FEP from $^{212}$Bi at 1621\,keV was used to test the performance of this cut.
The spectra observed in data and simulation are shown in Fig.~\ref{fig:psdcut} for the relevant energy range
together with the results of fits to the peaks, which were used to obtain survival fractions.
When choosing the $A/E$ cut for a 90\% survival fraction of events in the mostly single-site $^{208}$Tl DEP, the survival fraction of events in the mostly multi-site FEP at 1621\,keV from $^{212}$Bi decays becomes $(12.3 \pm 1.5)\%$ for data and $(10.5 \pm 1.4)\%$ for simulation in an analysis window of $(1621 \pm 5)$\,keV. 
The values agree within statistical uncertainties. 

The results of the study on $A/E$ validate the simulation and encourage current and future $0\nu\beta\beta$ decay experiments using HPGe detectors to evaluate the signal efficiency of their pulse-shape discrimination with {\it SolidStateDetectors.jl}. 


\section{Developments}
\label{sec:outlook}

This paper represents the status of {\it SolidStateDetectors.jl}
release\,v0.5. Substantial further developments are being incorporated
in release\,v1.0, which is planned for the fall of 2021. 
The upcoming features can be classified as related to
physics processes, tools to present the results or technical
upgrades to further improve the ease of usage and the 
execution speed.

The implementation of the detector physics will be improved:
\begin{itemize}
    \item In release\,v0.5, the drift of individual point charges are simulated and combined by superposition. 
    The interactions between them are ignored.
    Release\,v1.0 will have the option to drift charge clouds including the effects of self-repulsion and diffusion. 
    \item Already in release\,v0.5, the electric potential and field as well as the drift velocity fields can be calculated for undepleted regions of the detector.
    In release\,v1.0, the weighting potentials will also be calculable for undepleted detectors.
    \item  In release\,v0.5, the physics close to detector surfaces
    including the slow-down of charge carriers and trapping
    can be implemented by the user using virtual volumes.
    In release\,v1.0, a model will be implemented, where a charge cloud can be split and 
    charges which come too close to a surface will be stochastically trapped. The user
    will have to provide input parameters like probabilities and what "too close" means
    in a configuration file. 
    As these parameters are not really known the user might want to develop fits of predictions to data in order to determine them.     
\end{itemize}

A number of tools in release\,v1.0 will help to better understand
the detectors under study by providing information derived
from the simulation:
\begin{itemize}
    \item Prediction of the full-depletion voltage;
    \item Pulse-lengths for user defined pulse intervals, e.g.\
          from 5--95\% amplitude level;
    \item Isochrones, i.e.\ surfaces from which charge drifts 
          result in equal pulse-lengths;
    \item Improved visualization capabilities.
\end{itemize}

The technical improvements encompass:
\begin{itemize}
    \item Improved constructive solid geometry (CSG);
    \item Rotation, scaling and stretching of CSG primitives;
    \item Definition of surfaces;
    \item Distances to surfaces;   
    \item Support for fast calculation of fields on GPUs;
    \item Usage of multi-threading beyond field calculations;
    \item Extended documentation.
\end{itemize}

\section{Summary}
\label{sec:summary}

The open-source package {\it SolidStateDetectors.jl} has been introduced as a tool to simulate semiconductor detectors.
While special emphasis is given to germanium detectors, the package can also be adjusted to simulate the response of silicon, 
or any other kind of semiconductor detector.

The package was programmed in the language \julia{} with special care to ensure fast execution.
Constructive solid geometry is used to define a detector and its surroundings. 
The package can handle three-dimensional calculations as required for segmented detectors as well as the usage of symmetries for simpler devices.

An n-type segmented point-contact detector was used to demonstrate the capabilities of {\it SolidStateDetectors.jl} and to 
successfully verify the simulation by comparing predictions to data.
For rare event searches, accurate pulse-shape simulations can help to verify background identification techniques and provide guidance during detector manufacturing. 
A number of extensions for {\it SolidStateDetectors.jl} are planned for the near future. As an open-source package,
{\it SolidStateDetectors.jl} is also open for improvements and extensions from the community.

\bibliographystyle{JHEP}
\bibliography{SSDpaper}

\vskip 1cm
\centerline {\Large \bfseries Appendices}
\appendix
\section{Drift velocity models}
\label{app:drift}
The drift velocity models were implemented for germanium with standard parameters applicable for standard detectors~\cite{Hagemann-mt}. These models should, however, be directly applicable for silicon for the $\langle100\rangle$ and $\langle111\rangle$ axes, if the input parameters are adjusted. Further adjustments needed are provided as footnotes.

The drift of the charge carriers is governed by the electric field and the axes-dependent mobility tensors.
The mobility tensors for electrons and holes are not fully known for germanium, but different parametrizations exist to describe the experimental results.
As default, the drift velocity models from the AGATA Detector Library (ADL)~\cite{Bruyneel:2006764, Bruyneel:2016zih} have been implemented in {\it SolidStateDetectors.jl}.

Germanium has a face-centered cubic crystal structure.
When the electric field is aligned with one of the three principal crystallographic axes, $\langle100\rangle$, $\langle110\rangle$ or $\langle111\rangle$, the drift velocity vector is parallel to this axis.
The longitudinal velocity, $v_l$, describes the drift along these axes and is parametrized in the following way:
\begin{equation} \label{eq:vl}
    v_l(\mathcal{E}) = \frac{\mu_0 \mathcal{E}}{(1+(\mathcal{E}/\mathcal{E}_0)\,^\beta)^{1/\beta}} - \mu_n \mathcal{E} \text{\,,}
\end{equation}
where $\mathcal{E}$ is the electric field strength, while $\mu_0$, $\mathcal{E}_0$, $\beta$ and $\mu_n$ are parameters which were obtained by fits to measurements.
The values of these parameters for electrons and holes for the $\langle100\rangle$ and $\langle111\rangle$ axes are set in configuration files. This allows for their optimization for a given detector. 
From the drift velocities along these two axes, the drift velocity of electrons and holes in any direction can be derived.
For the calculations presented in this paper, default parameters~\cite{Bruyneel:2006764, Abt:2010ax} were used.
In the following, the coordinate system is chosen with the $z$ axis, i.e. the symmetry axis of cylindrical detectors, aligned with the $\langle001\rangle$ axis.

A default to calculate the general electron drift velocity~\cite{Mihailescu:2000jg} is implemented.
The conduction band in a germanium crystal reaches its minimal potential around the four equivalent $\langle111\rangle$ axes, called valleys\footnote{In silicon, the minimum of the conduction band is reached around the six equivalent $\langle100\rangle$ axes.}. The electron drift velocity vector, $\mathbf{v}_{e}$, can be written as
\begin{equation} \label{eq:eledrifta}
    \mathbf{v}_{e}(\bm{\mathcal{E}}) = \mathcal{A}(\mathcal{E})\sum_{j} \frac{n_j}{n} \frac{\gamma_j\bm{\mathcal{E}}_0}{\sqrt{\bm{\mathcal{E}}_0^\top\gamma_j\bm{\mathcal{E}}_0}} \text{\,,}
\end{equation}
where $\gamma_j$ is the tensor of inverted effective masses for the $j$th valley, $n_j/n$ is the fraction of carriers in the $j$th valley, $\bm{\mathcal{E}}_0$ is the normalized electric field vector and $\mathcal{A}(\mathcal{E})$ is a function of the magnitude of the electric field. The sum provides the direction of the electron drift. The $\gamma_j$ tensors are obtained by transforming the tensor of the inverted effective electron-masses
\begin{equation}
    \gamma_0 = 
    \begin{pmatrix}
        m_{t}^{-1} & 0 & 0 \\
        0 & m_{l}^{-1} & 0 \\
        0 & 0 & m_{t}^{-1}
    \end{pmatrix}
\end{equation}
from the coordinate system of the $j$th valley to the global $xyz$ coordinates, i.e.\;$\gamma_j=R_{j}^{-1}\gamma_0R_{j}$, where the rotation matrices are 
$R_{j}=R_{x}(\arccos{\sqrt{2/3}})\,R_{z}(\phi_{110}+j\pi/2)$. The angle between the $\langle110\rangle$ and the $y$ axis, $\phi_{110}$, is set by the user for a given setup.
The values $m_{l} = 1.64$ and $m_{t} = 0.0819$~\cite{Dexter:1956zz} are given in units of the electron rest mass\footnote{Different values for the effective electron masses, $m_l = 0.98$ and $m_t = 0.19$~\cite{Dexter:1956zz}, and different $R_{j}$ to transform $\gamma_0$ in silicon are used.}.

The deviation from an equal population of the valleys, $n_e/n=1/4$ in germanium\footnote{In silicon, $n_e/n = 1/6$.}, depends on the electric field as~\cite{Reik:1962zz}
\begin{equation} \label{eq:eldriftr}
    \frac{n_j}{n} = \mathcal{R}(\mathcal{E}) \left( \frac{(\bm{\mathcal{E}}_0^\top\gamma_j\bm{\mathcal{E}}_0)^{-1/2}}{\sum_i(\bm{\mathcal{E}}_0^\top\gamma_i\bm{\mathcal{E}}_0)^{-1/2}} - \frac{n_e}{n} \right) + \frac{n_e}{n} \text{\,,}
\end{equation}
where $\mathcal{R}(\mathcal{E})$ is a function of the electric field strength. 
If an electric field is applied along the $\langle100\rangle$ axis, the conduction bands are equally populated and the calculation of $\mathcal{A}(\mathcal{E})$ using Eq.\,(\ref{eq:eledrifta}) reduces to $\mathcal{A}(\mathcal{E}) = v^{\langle100\rangle}_{l,e}(\mathcal{E})/2.888$. Likewise, using Eq.\,(\ref{eq:eldriftr}) for $n_j/n$ in Eq.\,(\ref{eq:eledrifta}), $\mathcal{R}(\mathcal{E})$ can be calculated along the $\langle111\rangle$ axis to be 
$\mathcal{R}(\mathcal{E}) = -1.182\,v^{\langle111\rangle}_{l,e}(\mathcal{E})/\mathcal{A}(\mathcal{E})+3.161$\footnote{The expressions $\mathcal{A}(\mathcal{E}) = v^{\langle111\rangle}_{l,e}(\mathcal{E})/1.962$ and $\mathcal{R}(\mathcal{E}) = -3.925\,v^{\langle100\rangle}_{l,e}(\mathcal{E})/\mathcal{A}(\mathcal{E}) + 7.325$ result for silicon.}.
Here, $v^{\langle100\rangle}_{l,e}$ and $v^{\langle111\rangle}_{l,e}$ are the longitudinal electron-drift velocities along the respective axes.
The three numerical coefficients depend only on the effective mass values in $\gamma_0$.
Using these results together with Eq.\,(\ref{eq:vl}), the electron drift velocity vector can be calculated as a function of the electric field vector.

The default model to calculate the hole drift velocities in germanium~\cite{Bruyneel:2006764} assumes that only the heavy-hole valence band contributes to the hole mobility.
The following functions are used to calculate the components of the hole drift velocity depending on the mean wave vector of the heavy holes,
$\mathbf{k}_0 = (k_0, \theta_0, \phi_0)$, in the direction of the electric field in local spherical coordinates:
\begin{equation} \label{eq:holev}
\begin{aligned}
    v_r(\mathbf{k}_0) &= v^{\langle100\rangle}_{l,h}(\mathcal{E})\,[1 - \Lambda(k_0) (\sin(\theta_0)^4 \sin(2\phi_0)^2 + \sin(2\theta_0)^2)] \text{\,,} \\
    v_\theta(\mathbf{k}_0) &= v^{\langle100\rangle}_{l,h}(\mathcal{E})\,\Omega(k_0)\,[2 \sin(\theta_0)^3 \cos(\theta_0) \sin(2\phi_0)^2 + \sin(4\theta_0)] \text{\,,} \\
    v_\phi(\mathbf{k}_0) &= v^{\langle100\rangle}_{l,h}(\mathcal{E})\,\Omega(k_0)\,\sin(\theta_0)^3 \sin(4\phi_0) \text{\,.}
\end{aligned}
\end{equation}
The function $\Lambda(k_0)$ describes the different longitudinal velocities along the axes. The function $\Omega(k_0)$ governs the
deviation of the direction of the drift velocity vector from the direction of the electric field vector.
Their dependence on the mean wave number $k_0$ was obtained numerically~\cite{Bruyneel:2006764}:
\begin{equation}
\begin{aligned}
    \Lambda(k_0) &= -0.01322k_0 + 0.41145k_0^2 - 0.23567k_0^3 + 0.04077k_0^4 \text{\,,} \\
    \Omega(k_0) &= 0.006550k_0 - 0.19946k_0^2 + 0.09859k_0^3 -0.01559k_0^4 \text{\,,}
\end{aligned}
\end{equation}
where $k_0$ can be expressed using $v_{\mathrm{rel}} = v_{l,h}^{\langle111\rangle}(\mathcal{E})/v_{l,h}^{\langle100\rangle}(\mathcal{E})$ as
\begin{equation}
    k_0(v_{\mathrm{rel}}) = 9.2652 - 26.3467\,v_{\mathrm{rel}} + 29.6137\,v_{\mathrm{rel}}^2 - 12.3689\,v_{\mathrm{rel}}^3 \text{\,.}
\end{equation}

This parametrization, together with Eq.\,(\ref{eq:vl}) for $v_{l,h}^{\langle100\rangle}(\mathcal{E})$ and $v_{l,h}^{\langle111\rangle}(\mathcal{E})$, is used to calculate the components of the hole drift velocity vectors with Eq.\,(\ref{eq:holev}).
They are rotated from the local coordinates to the global $xyz$ coordinates as
\begin{equation}
    \mathbf{v}_{h} = 
    \begin{pmatrix}
        v_x \\ v_y \\ v_z
    \end{pmatrix}
    = R_z(\phi_0 +\frac{\pi}{4}+\phi_{110})\,R_{y}(\theta_0)\,
    \begin{pmatrix}
        v_{\theta} \\ v_{\phi} \\ v_{r}
    \end{pmatrix}
    \text{\,.}
\end{equation}
The predictions of this model have been verified~\cite{Bruyneel:2006764} with experimental data~\cite{Reggiani:1977hole}. 

There are indications that the description of mobilities may not be adequate for high-precision simulation of specific large volume detectors.
If at all possible, the user is advised to adjust the
mobility parameters for the detector design under study.


\section{Temperature dependence of drift velocities}
\label{app:temp}

The parameters for electrons and holes 
used to calculate the longitudinal drift velocities
for the $\langle100\rangle$ and $\langle111\rangle$ axes 
according to Eq.\,(\ref{eq:vl}) depend on the temperature, $T$.
The default parameters apply to the reference temperature, $T_0 = 78$\,K.
The simulation can be adjusted with four user defined functions 
$f_{l,e}^{\langle100\rangle}(T)$ and $f_{l,e}^{\langle111\rangle}(T)$, 
$f_{l,h}^{\langle100\rangle}(T)$ and $f_{l,h}^{\langle111\rangle}(T)$
to scale the longitudinal drift velocities as $v_l(\mathcal{E}, T) = f_l(T) v_l(\mathcal{E}, T_0)$ for electrons and holes, respectively.

Several functions $f_{l,e/h}(T)$ are predefined. A specific function
can be selected and its parameters can be defined in a dedicated configuration file. 
The scaling is defined for the two basic axes, $\langle100\rangle$ and $\langle111\rangle$, and is propagated
to all directions by the procedure described in Appendix~\ref{app:drift}.
The default is $f_{l,e/h}(T)=1$ for all directions and electrons and holes, i.e.\;the default values are used for all temperatures.

The most commonly used model for the $T$ dependence 
only considers the scattering of the electrons and holes off phonons in the crystal.
It suggests a form of 
$f_{l, e/h}(T) \sim T^{-3/2}$~\cite{Bardeen:1950zz}.
This is qualitatively supported by some measurements for electrons where an $f_{l,e}(T) \sim T^{-1.6}$ behavior was observed. However, for holes, an $f_{l,h}(T) \sim T^{-2.3}$ temperature dependence was observed~\cite{Prince:1953, Morin:1954}.
Other simple analytical models exist for higher temperatures which also include the temperature dependence of the saturation velocity in high fields~\cite{Omar:19871351}.
For some configuration, however, a Boltzmann-like temperature dependence of the drift time, $f_{l,e}^{-1}(T)=p_0+p_1e^{-p_2/T}$, was observed~\cite{Abt:2011gg}; these data were incompatible with a power law.
Any kind of functions $f_{l,e/h}(T)$ can be introduced with the recommendation that $f_{l,e/h}(T_0)=1$ as long as the default parameters~\cite{Bruyneel:2006764, Bruyneel:2016zih} for the drift models are used.

The effect of the temperature dependence of the drift velocities can be neglected for many qualitative studies.
However, the temperature dependence has to be taken into account for detailed comparisons of simulated pulses to data.


\end{document}